\documentstyle[twocolumn]{jpsj}

\def\runtitle{Hidden Order and Dimerization Transition in $S=2$ Chains}
\def\runauthor{M. Yamanaka, M. Oshikawa, and S. Miyashita}

\title{Hidden Order and Dimerization Transition in $S=2$ Chains}

\author{Masanori {\sc Yamanaka}\footnote{E-mail: yamanaka@kodama.issp.u-tokyo.ac.jp}
\footnote{Present address: Department of Applied Physics, University of Tokyo,
Bunkyo-ku, Tokyo 113.},
Masaki {\sc Oshikawa}$^{1,2,}$\footnote{E-mail: oshikawa@physics.ubc.ca} and
Seiji {\sc Miyashita}$^{3,}$\footnote{E-mail: miya@temp.ess.sci.osaka-u.ac.jp}} 

\inst{
Institute for Solid State Physics, University of Tokyo,
Minato-ku, Tokyo 106 \\
$^1$Physics Department, University of British Columbia,
6224 Agricultural Road, Vancouver, BC V6T 1Z1, Canada \\
$^2$Department of Applied Physics, University of Tokyo,
Bunkyo-ku, Tokyo 113 \\
$^3$Department of Earth and Space Science,
Osaka University, Machikaneyamacho, Toyonaka 560
}

\abst{
We study ground state properties of the $S=2$ quantum
antiferromagnetic chain with a bond alternation
\[
H = 
 \sum_{j} [ 1 + \delta (-1)^j ] 
     \mbox{\boldmath $S$}_{j} \cdot \mbox{\boldmath $S$}_{j+1}
\]
by a Quantum Monte Carlo calculation.
We find that the hidden $Z_2 \times Z_2$ symmetry is broken
for $0.3 < |\delta| < 0.5$ while it is unbroken in the other regions.
This confirms the successive dimerization transitions first
predicted by Affleck and Haldane.
Our result shows that these transitions can be understood
in terms of the hidden $Z_2 \times Z_2$ symmetry breaking, 
as was discussed using the Valence-Bond-Solid states.
Furthermore, we find that the behavior of the generalized string correlation
is qualitatively very similar to that in the Valence-Bond-Solid
states, including the location of zeroes as a function of the
angle parameter.
}

\kword{quantum spin chain, Haldane gap, bond-alternation,
hidden order, dimerization}

\begin{document}
\sloppy
\maketitle

%\section{Introduction}
Much effort has been devoted to examining
Haldane's conjecture\cite{Haldane:conj1,Haldane:conj2}
on the difference between quantum antiferromagnetic chains
with integer and half-odd-integer spins.
These studies also found several interesting phenomena besides
the original conjecture.
One of them is the successive transitions induced by a bond alternation.
Consider the Hamiltonian
\begin{equation}
  \label{eq:dimham}
  H =  \sum_{j} [ 1 + \delta (-1)^j ] 
        \mbox{\boldmath $S$}_{j} \cdot \mbox{\boldmath $S$}_{j+1},
\end{equation}
where $\delta$ represents the strength of the bond-alternation
($-1 \leq \delta \leq 1$).
The standard spin-$S$ Heisenberg antiferromagnetic
chain corresponds to $\delta=0$.
When $\delta= \pm 1$, this Hamiltonian reduces to that of
isolated dimers.
Affleck and Haldane~\cite{AffHal87} made an interesting prediction:
there are $2S$ critical transition points in
$-1 < \delta < 1$ for spin $S$.
(See also ref.~\citen{Affleck:LesHouches}.)
They mapped the model~(\ref{eq:dimham})
into an $O(3)$ non-linear sigma model with a topological term
defined by the Lagrangian
\begin{equation}
  {\cal L} = 
    \frac{1}{2g} (\partial_{\mu} \mbox{\boldmath $\varphi$})
                  (\partial^{\mu} \mbox{\boldmath $\varphi$})
    + \frac{\alpha}{8\pi} \epsilon^{\mu \nu} \mbox{\boldmath $\varphi$} \cdot
   ( \partial_{\mu} \mbox{\boldmath $\varphi$} \times
        \partial_{\nu} \mbox{\boldmath $\varphi$} ),
\label{eq:NLsigmalag}
\end{equation}
where {\boldmath  $\varphi$} is a three-component vector field with
the constraint {\boldmath $\varphi$}$^2 = 1$.
The parameters are related to those of the original model as follows:
\begin{equation}
  v=2S \sqrt{1-\delta^2}, \ \ 
% \;\;\;\;\;\;
  g = \frac{2}{S\sqrt{1-\delta^2}}, \ \ 
% \;\;\;\;\;\;
  \alpha = 2 \pi S (1+\delta),
\label{eq:NLSparam}
\end{equation}
where $v$ denotes the spin-wave velocity, which corresponds to the
``speed of light'' in the relativistic field theory.
Thus the topological angle $\alpha$ depends on the bond-alternation
parameter $\delta$. When $\delta$ is changed from $-1$ to $1$,
$\alpha$ changes from $ - 4 \pi S$ to $ 4 \pi S$.
It was argued that the above model has a massless (critical)
point for $\alpha \equiv \pi$ (mod $2\pi$),
namely $2 S$ critical points
are expected in model~(\ref{eq:dimham}) for $-1 < \delta < 1$.

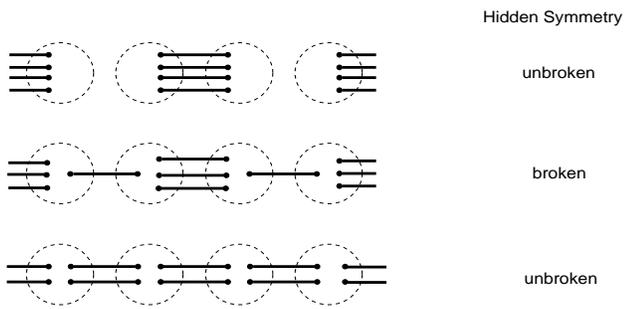
\begin{figure}[htbp]
%  \begin{center}
%%    \epsfxsize = \halftextwidth
%%    \epsfbox{s2vbsdim.eps}
%    \epsfile{file=s2vbsdim.eps,width=60mm}
%    \leavevmode
%  \end{center}
  \caption{The VBS picture of the $S=2$
successive dimerization transitions.
Each dot represents a spin $1/2$.
The bonds represent singlet pairs (``valence bonds''),
and the broken circle indicates the symmetrization of
four spin-$1/2$'s to form a spin $2$.
The number of valence bonds between the neighboring sites
changes as the dimerization proceeds.
The hidden $Z_2 \times Z_2$ symmetry is broken only
in the intermediate state.}
\label{fig:s2vbsdim}
\end{figure}

There is another (more intuitive) argument~\cite{AffHal:unp}
based on a class of exactly solvable spin chains --
Valence-Bond-Solid (VBS) models.\cite{AKLT}
We can introduce VBS states with various degrees of dimerization.
For $S=2$, each spin can be constructed
by symmetrization of four spin-$1/2$'s
and we can consider VBS-type states as shown in fig.~\ref{fig:s2vbsdim}.
In general, $2S +1$ states can be constructed for spin $S$.
If we assume that the ground state of the Hamiltonian~(\ref{eq:dimham})
belongs to the same phase as one of these states, and that
there are phase transitions between them, there should be $2S$ phase
transitions as expected in the field-theory argument.

However, both approaches are based on some assumptions or approximations
and thus the prediction should be verified by, for example, a numerical
calculation.
In addition, the concept of the order parameter is lacking in these
arguments.
How can we distinguish the phases which correspond to
``different numbers of valence bonds''?
For $S=1$, den Nijs and Rommelse~\cite{dNR89}
found that a kind of non-local
order parameter characterizes the Haldane gap phase.
(See also ref.~\citen{Hal:sto}.)
Kennedy and Tasaki~\cite{KennedyTasaki:hid}
revealed that this den Nijs-Rommelse string
order parameter measures a hidden $Z_2 \times Z_2$ symmetry breaking.
One of us extended this idea in ref.~\citen{MOVBS} to general
integer spin, and found that the VBS states can be (partly) distinguished by
the hidden $Z_2 \times Z_2$ symmetry breaking.
Namely, it was found that the symmetry is broken (unbroken)
when the number of valence bonds is odd (even).
Thus the dimerization transitions can be understood as
hidden $Z_2 \times Z_2$ symmetry breaking, if the VBS picture is correct.
(See fig.~\ref{fig:s2vbsdim}.)

However, the hidden $Z_2 \times Z_2$ symmetry is not sufficient for
full characterization of all possible VBS states.
Also it cannot be applied to half-odd-integer spin chains.
As an attempt to overcome these problems, a generalized string order
was proposed.\cite{MOVBS}
In a dimerized state, the generalized string order depends on
how the limit is taken.
Here we choose the definition
\begin{equation}
\label{eq:genstrdim}
 O^{\alpha}_{\rm string}(\theta) = \lim_{|k-j| \rightarrow \infty} 
 \left\langle S^z_{2j+1} 
    \exp{({\rm i} \theta \sum_{l=2j+1}^{2k} S^z_l )} S^z_{2k+1}
 \right\rangle .
\end{equation}
This reduces to the N\'{e}el order for $\theta =0$, and to the
den Nijs-Rommelse string order for $\theta=\pi$, which is related
to the hidden $Z_2 \times Z_2$ symmetry breaking for integer spin.

It was found\cite{MOVBS} that the generalized string
order parameter in the ($n,m$)-VBS state
(with $n$ valence bonds between sites $2j$ and $2j+1$
and $m$ between $2j+1$ and $2j+2$)
is given by a polynomial of $z = {\rm e}^{{\rm i} \theta}$ as
\begin{equation}
  O^{\alpha}_{\rm string}(\theta)
= \left| f_{n, m}({\rm e}^{{\rm i} \theta}) \right|^2 ,
\label{eq:vbsgenstr} 
\end{equation}
where $f$ is given by
\begin{equation}
\label{eq:resfnm}
 f_{n, m}(z) = \frac{n+m+2}{2(n+2)(n+1)}
                  \sum_{k=0}^n (2k-n) z^k .
\end{equation}
We note that $O^{\alpha}_{\rm string}$ is not the same
for ($n,m$)- and ($m,n$)-VBS states.
The order parameter defined as eq.~(\ref{eq:genstrdim})
in the ($n,m$)-VBS state is equal to an order parameter
in the ($m,n$)-VBS state with a slightly different definition.
It was proved\cite{MOVBS}
that the $n$-th order polynomial $f_{n,m}$ has
$n$ simple roots on the unit circle. Thus the generalized string
order has $n$ zeroes as a function of $\theta$ in $0 \leq \theta < 2 \pi$.
It was conjectured that this property is generic for the partially dimerized
phases.
We also note that
$O^{\alpha}_{\rm string}(\theta) = O^{\alpha}_{\rm string}(2\pi - \theta)$
and it is always real.

For $S=1$, the presence of the critical transition point has been
numerically established by several
authors.\cite{KatoTanaka,Yamamoto:S=1dim,Totsukaetal:S=1dim}
Recently Yajima and Takahashi\cite{YajimaTakahashi}
studied the $S=3/2$ case and confirmed the presence of the transitions.
They also measured the generalized string order parameter at $\theta=\pi$
(this is not related to the hidden symmetry for half-odd-integer spin)
and found the behavior consistent with the above VBS picture.
For $S=2$, the Heisenberg point ($\delta = 0$) was studied by several
authors\cite{Hatsugai:S=2str,HatanoSuzuki93:S=2,Deiszetal93:spin,Meshkov93:spin,Nishiyamaetal95:S=2,Joli,OYM,Yamamoto:S=2corr}
and the behavior of the string order parameter was found to be consistent with
the VBS picture.
Tonegawa numerically calculated the energy gap
of the model (\ref{eq:dimham})
with the next-nearest-neighbour couplings.\cite{Tone:S=2}

In this paper, we study the ground state properties of the $S=2$ chain with
the Hamiltonian~(\ref{eq:dimham}) more thoroughly
by the Quantum Monte Carlo method,
with particular emphasis on the hidden order and the hidden symmetry breaking.

%\section{Method}
We performed a world-line
Quantum Monte Carlo calculation~\cite{Suzuki76:Trotter} using the
Lie-Trotter-Suzuki product formula
with checker-board decomposition.~\cite{Hirshetal:checkerboard}
That is, we made an approximation to the partition function $Z$
for temperature $T$ as
%\begin{equation}
%\label{eq:decomposed}
$
Z_n = \mbox{Tr}[( {\rm e}^{-{H_A}/{(nT)}} {\rm e}^{-{H_B}/{(nT)}})^n].
$
%\end{equation}
Here we choose
$H_A = \sum_{j = {\rm odd}}  V_j$ ,
$H_B = \sum_{j = {\rm even}} V_j$ and
$V_j = [ 1 + \delta (-1)^j ]
   \mbox{\boldmath S}_{j} \cdot \mbox{\boldmath S}_{j+1} $.
The approximate partition function
$Z_n$ approaches the true partition function $Z$ as
$n \rightarrow \infty$. 
We chose a heat-bath algorithm.
While we prepared global flips along the chain direction, 
we did not use global flips along the Trotter direction and
restricted the calculation to the $\sum S^z = 0$ subspace,
in order to study the ground state.
We used the periodic boundary condition.
For most of the calculation we took $T=0.04$ and $n=96$,
which turned out to be sufficiently close to the
extrapolation $T \rightarrow 0$ and
$n \rightarrow \infty$.

%\section{Results}
We measured the generalized string correlation~(\ref{eq:genstrdim})
at $\theta = n \pi /20$ ($n=0,1, \ldots, 20$).
We assumed the 
$L$ dependence of the correlation function at distance $L/2$ to be
\begin{equation}
 \left\langle S^z_{0} 
    \exp{({\rm i} \theta \sum_{l=0}^{L/2-1} S^z_l )} S^z_{L/2}
 \right\rangle
\simeq  A + B \exp{( - C L)} .
\label{eq:fitting}
\end{equation}
We made a least-square fit of our data for system size
$L(=40$, 60, 80) with the above formula and
checked the stability of the extrapolation for different
choices of the system size ($L=40$, 50, 60, 70, 80).
%\subsection{The Heisenberg point}

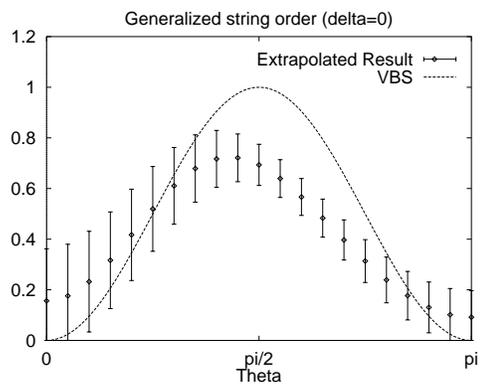
\begin{figure}[htbp]
%  \begin{center}
%%    \epsfxsize = \halftextwidth
%%    \epsfbox{dltm00genstr.eps}
%    \epsfile{file=dltm00genstr.eps,width=90mm}
%    \leavevmode
%  \end{center}
  \caption{The extrapolated generalized string order at $\delta=0$
as a function of $\theta$.
The exact result in the VBS state is also shown.
Their behaviors are quite similar, and both have $\theta=0,\pi$
as zeroes.
The relatively large error is presumably due to 
the large correlation length compared to the cases for 
$\delta=0.4$ and $0.6$
}
\label{fig:dlt0genstr}
\end{figure}

{\it The Heisenberg point:}
The generalized string order $A$ obtained by the fitting~(\ref{eq:fitting})
and the result~(\ref{eq:vbsgenstr}) for the $S=2$ VBS state
are shown in fig.~\ref{fig:dlt0genstr}.
Their qualitative behaviors are very similar.
In particular, they have zeroes at $\theta =0, \pi$.
The former corresponds to vanishing N\'{e}el order and
the latter means that the hidden $Z_2 \times Z_2$ symmetry is unbroken.
These results are in agreement with the VBS picture,
and also with previous numerical
calculations\cite{HatanoSuzuki93:S=2,Deiszetal93:spin,Meshkov93:spin,Nishiyamaetal95:S=2,Joli,OYM,Yamamoto:S=2corr}
at the Heisenberg point.
The generalized string order
has its maximum value at a point near $\theta = \pi /2$.
The dependence of the generalized string order on $\theta$ was
studied in a short chain by Hatsugai.\cite{Hatsugai:S=2str}
The long-range order at $\theta = \pi /2$ was confirmed
numerically in refs.~\citen{Nishiyamaetal95:S=2} and~\citen{Joli}.
Here we emphasize that the generalized string order is
non-vanishing for $\theta \neq 0, \pi$, 
in agreement with the result~(\ref{eq:vbsgenstr}) for the VBS state.
We observed a similar result at $\delta=0.05$.

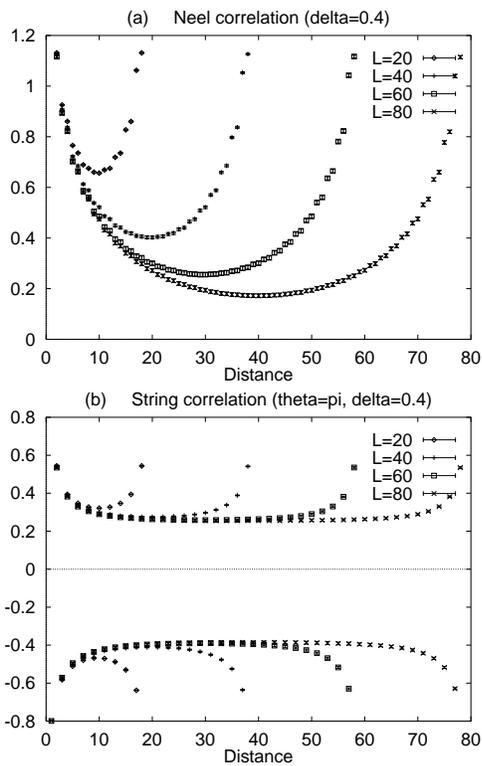
\begin{figure}[htbp]
%  \begin{center}
%%    \epsfxsize = \halftextwidth
%%    \epsfbox{dltm04neel.eps}
%%    \epsfxsize = \halftextwidth
%%    \epsfbox{dltm04strpi.eps}
%    \epsfile{file=dltm04neel.eps,width=90mm}
%    \epsfile{file=dltm04strpi.eps,width=90mm}
%    \leavevmode
%  \end{center}
  \caption{(a) The N\'{e}el and (b) the den Nijs-Rommelse string correlations
at $\delta=0.4$ for system sizes $L=20,40,60$ and $80$. 
The den Nijs-Rommelse string correlation is long-ranged,
while the N\'{e}el correlation decays exponentially.}
\label{fig:dlt04corr}
\end{figure}

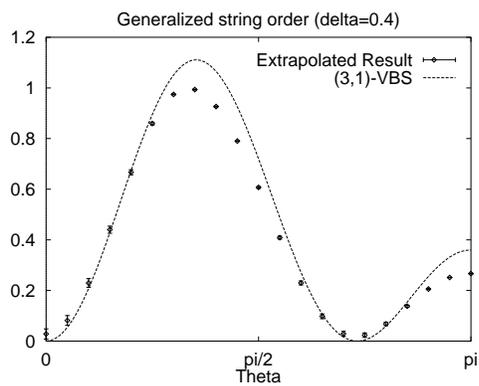
\begin{figure}[htbp]
%  \begin{center}
%%    \epsfxsize = \halftextwidth
%%    \epsfbox{dltm04genstr.eps}
%    \epsfile{file=dltm04genstr.eps,width=90mm}
%    \leavevmode
%  \end{center}
  \caption{The extrapolated generalized string order at $\delta=0.4$
as a function of $\theta$.
The exact result for the partially dimerized VBS state is also shown.
Their behaviors are quite similar, including the
location of the zero.}
\label{fig:dlt04genstr}
\end{figure}

%\subsection{The intermediate phase}
{\it The intermediate phase:}
We show the N\'{e}el and 
the den Nijs-Rommelse string correlation functions at $\delta = 0.4$
in fig.~\ref{fig:dlt04corr}.
The N\'{e}el correlation decays faster than that at the Heisenberg point.
Due to the bond-alternation, 
the den Nijs-Rommelse string correlation has an oscillating part.
Besides the oscillating part,
it seems to have a long-range order.
This means that the hidden $Z_2 \times Z_2$ symmetry
is broken.
We show the extrapolated results in fig.~\ref{fig:dlt04genstr}
together with the result~(\ref{eq:vbsgenstr})
for the partially dimerized VBS state.
The result again agrees (qualitatively) with that of the VBS state.
At $\theta=\pi$ (den Nijs-Rommelse string order) there is no longer zero
and instead we see another zero in $0 < \theta < \pi$.
By symmetry, there are three zeroes in $0 \leq \theta < 2\pi$
in agreement with the conjecture in ref.~\citen{MOVBS}.
It is somewhat surprising that
not only the number of zeroes, but also
the location of the zeroes in $\theta$
seems to coincide with the VBS result.
For $\delta=0.3$ and $0.5$ we obtained similar results
to that for $\delta=0.4$.

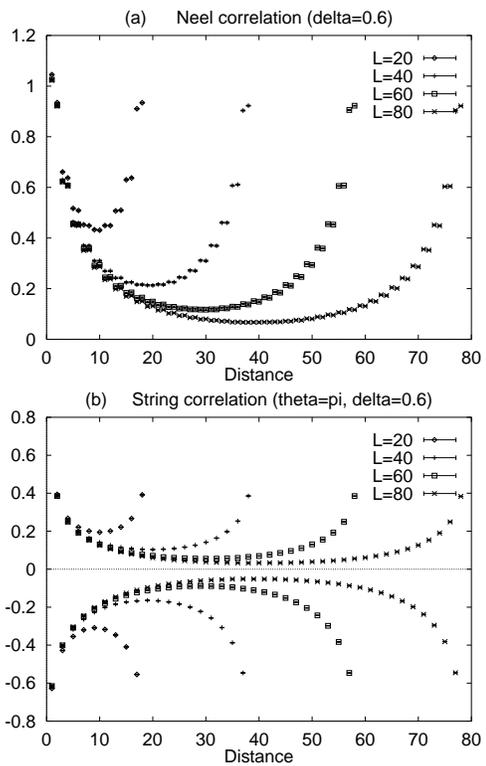
\begin{figure}[htbp]
%  \begin{center}
%%    \epsfxsize = \halftextwidth
%%    \epsfbox{dltm06neel.eps}
%%    \epsfxsize = \halftextwidth
%%    \epsfbox{dltm06strpi.eps}
%    \epsfile{file=dltm06neel.eps,width=90mm}
%    \epsfile{file=dltm06strpi.eps,width=90mm}
%    \leavevmode
%  \end{center}
  \caption{(a) The N\'{e}el and (b) the den Nijs-Rommelse string correlations
at $\delta=0.6$ for system sizes $L=20,40,60$ and $80$. 
Both correlation functions decay to zero.}
\label{fig:dlt06corr}
\end{figure}

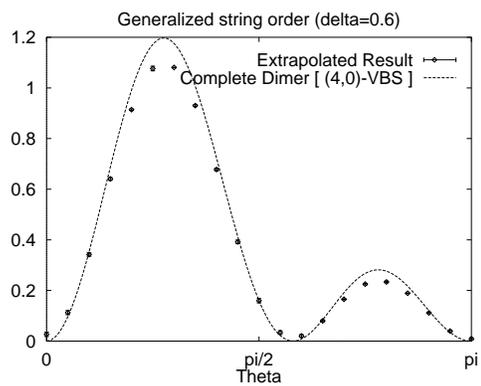
\begin{figure}[htbp]
%  \begin{center}
%%    \epsfxsize = \halftextwidth
%%    \epsfbox{dltm06genstr.eps}
%    \epsfile{file=dltm06genstr.eps,width=90mm}
%    \leavevmode
%  \end{center}
  \caption{The extrapolated generalized string order at $\delta=0.6$
as a function of $\theta$.
The exact result in the completely dimerized state is also shown.
Their behaviors are quite similar, including the
location of the zero.}
\label{fig:dlt06genstr}
\end{figure}

%\subsection{The dimerized phase}
{\it The dimerized phase:}
Next we consider $\delta = 0.6$.
As  shown in fig.~\ref{fig:dlt06corr},
both the N\'{e}el correlation and the den Nijs-Rommelse string
correlation decay to zero.
The latter means that
the hidden $Z_2 \times Z_2$ symmetry is unbroken.
We show the extrapolated results in fig.~\ref{fig:dlt06genstr},
together with the result~(\ref{eq:vbsgenstr})
for the completely dimerized state.
Their behaviors agree well.
In particular, the locations of zeroes seem to agree again.
For $\delta=0.7$ and $0.8$, we observed similar behaviors
to that for $\delta=0.6$
(but more similar to the result for the completely dimerized state,
as expected).

%\section{Summary and discussion}
To summarize, we measured the generalized string correlation
in the $S=2$ spin chain with a bond-alternation~(\ref{eq:dimham})
by the Quantum Monte Carlo method.

We observed that there are three phases in $0 < \delta < 1$, 
confirming the prediction by Affleck and Haldane.
The hidden $Z_2 \times Z_2$ symmetry is broken only in the
intermediate phase. This is consistent with the VBS picture of the
transitions.
Hence the transitions can be understood as
hidden $Z_2 \times Z_2$ symmetry breaking.
From this argument, it would be natural to expect that the transitions
belong to the same $c=1$ universality class as in the $S=1$ case.
This argument is again consistent with the field-theory prediction by
Affleck and Haldane.

Furthermore, we found that the behaviors of the generalized string order in
those phases are quite similar to those in the VBS-type state.
In particular, the locations of the zeroes seem to agree.
It might be possible to determine the underlying mechanism for this phenomenon.
(For example, the zeroes may be related to some unknown hidden symmetries.)

Our result shows that the region
$0 \leq \delta \leq 0.05$ belongs to the $S=2$ Haldane phase,
$0.3 \leq \delta \leq 0.5$ to the intermediate phase,
and $0.6 \leq \delta \leq 1$ to the dimerized phase.
Thus the transition points should be located
in the regions $0.05 < \delta < 0.3$ and $0.5 < \delta < 0.6$.
However, in these regions, the correlation length seems to grow
and hence we could not make reliable extrapolation and could not
estimate the transition points more precisely.

From eq.~(\ref{eq:NLSparam}) we can obtain the field-theory
prediction for the transition points:
$\delta = 0.25$ and $\delta = 0.75$.
The latter is clearly inconsistent with our result $ 0.5 < \delta < 0.6$.
(The value was not consistent with the numerical
result for $S=1$\cite{KatoTanaka,Yamamoto:S=1dim,Totsukaetal:S=1dim}
and for $S=3/2$.\cite{YajimaTakahashi})
In general, we cannot expect that the field theoretical argument gives the
quantitatively precise location of the critical point. 
However, the qualitative prediction of the occurrence of the
phase transition and the description of the critical behavior
are reliable. 

\section*{Acknowledgments}
The authors thank Hal Tasaki for useful discussions. 
One of the authors (M.Y.) acknowledges financial support
from the JSPS Research Fellowships for Young Scientists.
Numerical calculations were partially carried out
using the facilities of the Supercomputer Center,
Institute for Solid State Physics, University of Tokyo. 

%\newpage
\bibliographystyle{prsty}

%\bibliography{oshi}

\begin{thebibliography}{10}

\bibitem{Haldane:conj1}
F.D.M. Haldane: Phys. Lett. {\bf 93A} (1983) 464.

\bibitem{Haldane:conj2}
F.D.M. Haldane: Phys. Rev. Lett. {\bf 50} (1983) 1153.

\bibitem{AffHal87}
I. Affleck and F.D.M. Haldane: Phys. Rev. B {\bf 36} (1987) 5291.

\bibitem{Affleck:LesHouches}
I. Affleck:  in {\em Fields, Strings and Critical Phenomena}, {\em Les Houches,
  Session XLIX}, edited by E. Brezin and J. Zinn-Justin (North-Holland,
  Amsterdam, 1988).

\bibitem{AffHal:unp}
I. Affleck and F.D.M. Haldane: unpublished.

\bibitem{AKLT}
I. Affleck, T. Kennedy, E. Lieb, and H. Tasaki:
Commun. Math. Phys. {\bf 115} (1988) 477.

\bibitem{dNR89}
M. den Nijs and K. Rommelse: Phys. Rev. B {\bf 40} (1989) 4709.

\bibitem{Hal:sto}
H. Tasaki: Phys. Rev. Lett. {\bf 66} (1991) 798.

\bibitem{KennedyTasaki:hid}
T. Kennedy and H. Tasaki: Commun. Math. Phys. {\bf 147} (1992) 431.

\bibitem{MOVBS}
M. Oshikawa: J. Phys. Condens. Matter {\bf 4} (1992) 7469.

\bibitem{KatoTanaka}
Y. Kato and A. Tanaka: J. Phys. Soc. Jpn. {\bf 63} (1994) 1277.

\bibitem{Yamamoto:S=1dim}
S. Yamamoto: J. Phys. Soc. Jpn. {\bf 63} (1994) 4327. 

\bibitem{Totsukaetal:S=1dim}
K. Totsuka, Y. Nishiyama, N. Hatano, and M. Suzuki:
J. Phys. Condens. Matter {\bf 7} (1995) 4895.

\bibitem{YajimaTakahashi}
M. Yajima and M. Takahashi: J. Phys. Soc. Jpn. {\bf 65} (1996) 39. 

\bibitem{Hatsugai:S=2str}
Y. Hatsugai: J. Phys. Soc. Jpn. {\bf 61} (1992) 3856. 

\bibitem{HatanoSuzuki93:S=2}
N. Hatano and M. Suzuki: J. Phys. Soc. Jpn. {\bf 62} (1993) 1346. 

\bibitem{Deiszetal93:spin}
J. Deisz, M. Jarrell, and D.J. Cox: Phys. Rev. B {\bf 48} (1993) 10227.

\bibitem{Meshkov93:spin}
S.V. Meshkov: Phys. Rev. B {\bf 48} (1993) 6167.

\bibitem{Nishiyamaetal95:S=2}
Y. Nishiyama, K. Totsuka, N. Hatano, and M. Suzuki:
J. Phys. Soc. Jpn. {\bf 64} (1995) 414.

\bibitem{Joli}
U. Schollw\"{o}ck and T. Jolicoeur: Euro. Phys. Lett. {\bf 30} (1995) 493.

\bibitem{OYM}
M. Oshikawa, M. Yamanaka, and S. Miyashita: unpublished.

\bibitem{Yamamoto:S=2corr}
S. Yamamoto: Phys. Lett. A, {\bf 213} (1996) 102.

\bibitem{Tone:S=2}
T. Tonegawa: unpublished.

\bibitem{Suzuki76:Trotter}
M. Suzuki: Prog. Theor. Phys. {\bf 56} (1976) 1454.

\bibitem{Hirshetal:checkerboard}
J.E. Hirsch, R.L. Sugar, D.J. Scalapino, and R. Blankenbecler:
Phys. Rev. B {\bf 26} (1982) 5033.

\end{thebibliography}

%\newpage
\vspace{-7mm}
\vspace{-7mm}
\vspace{-7mm}
\vspace{-7mm}
\vspace{-7mm}
\end{document}